\documentclass[aps, pra, reprint, superscriptaddress, longbibliography]{revtex4-2}

\pdfoutput=1

\usepackage{amsmath}
\usepackage{amsfonts}
\usepackage{amssymb}
\usepackage{amstext}   
\usepackage{amsthm}
\usepackage{dsfont}
\usepackage{hyperref}
\usepackage{graphicx}
\usepackage{mathtools}
\usepackage{mathdots}
\usepackage{braket}
\usepackage[capitalize]{cleveref}
\usepackage[notransparent]{svg}
\usepackage{xcolor}
\usepackage{physics}
\usepackage{graphicx}
\hypersetup{
    colorlinks=true,
    linkcolor=blue,
    filecolor=red,      
    urlcolor=red,
    citecolor=red
}

\newcommand{\avg}[1]{\ensuremath{\left\langle#1\right\rangle}}
\newcommand{\mc}[1]{\mathcal{#1}}
\newcommand{\wt}[1]{\widetilde{#1}}

\newcommand{\qmaddress}{\affiliation{Quantum Motion, 9 Sterling Way, London N7 9HJ, United Kingdom}}

\theoremstyle{main}

\theoremstyle{plain}

\theoremstyle{definition}

\theoremstyle{remark}

\newtheorem{statement}{Statement}

\definecolor{emerald}{rgb}{0.07, 0.53, 0.03}

\renewcommand{\tr}{\mathrm{tr}}
\newcommand{\expect}[1]{{\mathbb{E}[#1]}}
\renewcommand{\var}[1]{{\mathrm{Var}[#1]}}

\newcommand{\mse}[1]{{\mathrm{MSE}[#1]}}

\begin{document}

\title{Biased Estimator Channels for Classical Shadows}

\author{Zhenyu Cai}
\thanks{alphabetical author list}
\affiliation{Department of Materials, University of Oxford, Parks Road, Oxford OX1 3PH, United Kingdom} 
\qmaddress

\author{Adrian Chapman}
\email{adrian.chapman@materials.ox.ac.uk}
\affiliation{Department of Materials, University of Oxford, Parks Road, Oxford OX1 3PH, United Kingdom} 

\author{Hamza Jnane}
\affiliation{Department of Materials, University of Oxford, Parks Road, Oxford OX1 3PH, United Kingdom} 
\qmaddress

\author{Bálint Koczor}
\email{koczor@maths.ox.ac.uk}
\affiliation{Mathematical Institute, University of Oxford, Woodstock Road, Oxford OX2 6GG, United Kingdom}
\affiliation{Department of Materials, University of Oxford, Parks Road, Oxford OX1 3PH, United Kingdom} 
\qmaddress

\begin{abstract}
Extracting classical information from quantum systems is of fundamental importance, and classical shadows allow us to extract a large amount of information using relatively few measurements. 
Conventional shadow estimators are unbiased and thus agree with the true mean in expectation.
In this work, we consider a biased scheme, intentionally introducing a bias in the expectation value by rescaling the conventional classical shadows estimators to reduce the error in the finite-sample regime. The approach is straightforward to implement and requires no quantum resources. We analytically prove average case as well as worst- and best-case scenarios, and rigorously prove that it is, in principle, always worth biasing the estimators. We illustrate our approach in a quantum simulation task of a $12$-qubit spin-ring problem  and demonstrate how estimating expected values of non-local perturbations can be significantly more 
efficient using our biased scheme.
\end{abstract}

\maketitle

\section{Introduction}
We are experiencing rapid progress in the development of quantum hardware but also in theoretical  advances~\cite{aruteQuantumSupremacyUsing2019, wuStrongQuantumComputational2021, zhongPhaseProgrammableGaussianBoson2021, bluvstein_logical_2023, kim_evidence_2023, acharya_suppressing_2023}.
Any quantum computational scheme needs to extract classical information from a quantum device.
However, this requires multiple repetitions of the experiment due to fundamental limitations posed by quantum mechanics.
Each observation of the system collapses its quantum state, preventing one from extracting further information.
For this reason, one must extract classical information through the use of statistical estimators~\cite{sugiyama2013precision, guta2020fast, meyer2023quantum, elben2023randomized}.
It is thus an exciting and fundamentally important challenge to extract classical information, for example, expected values of observables, such that the statistical uncertainty (i.e., shot noise due to having access to only finite samples or circuit repetitions) is minimised.
Classical shadows~\cite{huang2020predicting} allow one to predict many such properties of quantum states from very few samples (circuit repetitions) and the statistical uncertainty due to shot noise can be rigorously bounded.
The approach yields unbiased estimators, i.e., estimators whose expected value agrees with the true value, e.g., the true expected value of an observable.

In the present work, we explore the possibility of intentionally introducing a small bias
into the estimators, i.e., for a fixed such bias, even an infinite number of samples would not yield the true expected value, but in return the statistical uncertainties are significantly reduced in the finite-sample regime. A significant practical advantage of our biasing scheme is that it is implemented completely in
classical post processing: one collects a number of samples as classical shadows using a quantum computer,
and in post-processing predicts many properties of the quantum state. Our approach 
only slightly modifies  this prediction stage whereby the mean estimators are simply scaled down by a factor
quantified by a bias parameter $\varepsilon$. 
We choose this parameter to depend on the number of samples in the experiment such that the resulting estimator remains consistent, converging to the true value in the infinite-sample limit.
The approach is thus more general than the standard, unbiased shadow techniques~\cite{huang2020predicting,huang2021efficient, zhao2021fermionic, wan2022matchgate, chen2021robust, nguyen2022optimizing, ferrie2018maximum, Akhtar2023scalableflexible, bu_classical_2024, Koh2022classicalshadows} which are then contained as a special case at $\varepsilon =0$.

We comprehensively and rigorously characterise the performance of our biased scheme and find,
somewhat surprisingly,
that biasing our estimators is always worthwhile assuming that an optimal bias parameter $\varepsilon$
that is specific for the particular estimator is known. 
After briefly recapitulating classical shadows, we start by mathematically deriving the average-case
gain of biased shadow tomography when the aim is to predict local density matrices.
We then analytically characterise 
both the worst, and best-case scenarios of our approach when the aim is to predict expected values of observables,
and we provide explicit expressions for the optimal bias parameter $\varepsilon$ showing it
depends only on the theoretical mean value.
We then argue that the biased scheme is not specific to classical shadows but can 
generally be applied to any estimation scheme, e.g., by directly estimating Pauli expected values.

\section{Classical shadows}

A classical shadow is a description of a quantum state that can be classically efficiently stored and manipulated, enabling one to bypass the computationally hard task of reconstructing the full density matrix.
To construct a classical shadow for an $n$-qubit quantum state $\rho$, we repeat the following evolution-measurement process.
We sample a random unitary $U_i$ from a suitable distribution $\mc{U}$ (Pauli and Clifford distributions are typical), apply this unitary to $\rho$, and measure the resulting state $U_i \rho U_i^{\dagger}$ in the computational basis, yielding a bitstring $\mathbf{b} \in \{0, 1\}^{n}$.
We store the index of the unitary and the measurement outcome as a composite index $\ell = (i, \mathbf{b})$.

One considers the \emph{process channel} $\mc{M}$
as the average over the previously fixed distribution $\mc{U}$ as
\begin{align}
    \mc{M}(\rho) = \sum_{\mathbf{b} \in \{0, 1\}^n} \int_{\mc{U}} dU \bra{\mathbf{b}} U^{\dagger} \rho U \ket{\mathbf{b}} U \ketbra{\mathbf{b}}{\mathbf{b}} U^{\dagger} \mathrm{.}
\end{align}
Each composite index $\ell$ then identifies a classical snapshot $\hat{\rho}_{\ell}$ such that, 
\begin{align} \label{eq:estimator_channel}
    \hat{\rho}_{\ell} = \wt{\mc{M}}(U_i \ketbra{\mathbf{b}}{\mathbf{b}} U_{i}^{\dagger})
\end{align}
which is an unbiased estimator of $\rho$.
Here, $\wt{\mc{M}}$ is the \emph{estimator channel}, which is implicitly defined by this equation.
In practice, one repeats the above procedure $N_s$ times, generating the \emph{classical shadow} of $\rho$
as the collection $S(\rho, N_s) = \{\hat{\rho}_1, \dots, \hat{\rho}_{N_s}\}$.
When discussing averages over $\ell$, we drop the corresponding dependence on $\hat{\rho}$ and understand this estimator to be a function of $\ell$.

\begin{figure}[t!]
    \centering
    \includegraphics[width=0.45\textwidth]{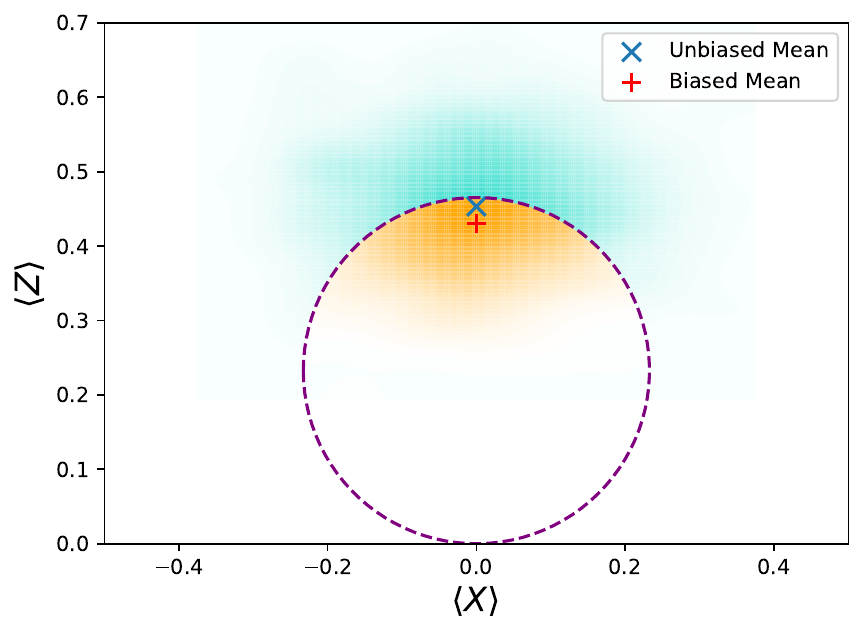}
    \caption{A geometric illustration of variance-bias tradeoff, displayed via the density of \emph{unbiased} Pauli-$X$ and -$Z$ estimates, where each estimate is obtained as the average over $N_s = 100$ samples.
    The exterior of the dashed circle corresponds to the region where biasing would hypothetically bring the estimate closer to the true expectation value (with $\avg{X} = 0$ and $\avg{Z} = 0.45\bar{3}$) for a biasing parameter of $\varepsilon = 0.1$.
    That is, the color of the density cloud represents the \emph{sign} of the change in loss upon biasing.
    The radius of the dashed circle is increasing with $\varepsilon$, but this is compensated by the magnitude of the decrease in loss so that biasing improves the estimate for $\varepsilon \leq \varepsilon_{\mathrm{min}}$.
    See details in \cref{app:single_qubit_error}.
    }
    \label{fig:densityplot}
\end{figure}

\section{Main Result}
\subsection{Biased shadow estimators}
A channel $\widetilde{\mathcal{M}}$ that yields an unbiased estimator 
for the density matrix satisfies the condition $(\wt{\mc{M}} \circ \mc{M})[\rho] = \rho$ for all states $\rho$,
and thus guarantees $\wt{\mc{M}} = \mc{M}^{-1}$ as
\begin{align}
	\rho = \mathds{E}_{\mc{U}, \mathbf{b}}[\hat{\rho}] = (\wt{\mc{M}} \circ \mc{M})[\rho], ~~ \forall \rho
	\label{eq:mtom}
\end{align}
However, in the present work we focus on constructing a biased estimator 
which does not necessarily satisfy the above property
but in return allows us to reduce the variance of the estimator.

To simplify our presentation, we illustrate our results on the simple uniform ensemble over $n$-qubit product Clifford rotations
$\mc{U} = \mc{C}^{\times n}_1$.
This is equivalent to uniformly sampling a local Pauli basis in which to measure each qubit
and thus $\hat{\rho}_{\ell}$ can be decomposed as a tensor product \cite{huang2020predicting},
\begin{align}
    \hat{\rho}_{\ell} &= \bigotimes_{j=1}^n \hat{\rho}_{\ell}^{(j)} = \bigotimes_{j=1}^n\Big[ 3 (U^{(j)}_{i})^{\dagger} \vert b^{(j)}\rangle \langle b^{(j)} \vert  U^{(j)}_{i} - \openone \Big], 
\end{align}
where the superscript $(j)$ indicates that we consider the $j^{th}$ term in the tensor product decomposition. We can then rewrite $\hat{\rho}_{\ell}$ and define $\wt{\mc{M}}_{\mathrm{local}}$ as
\begin{align} \label{eq:snapshot}
    \hat{\rho}_{\ell} &\equiv \bigotimes_{j=1}^n \wt{\mc{M}}_{\mathrm{local}}[(U^{(j)}_{i})^{\dagger} \vert b^{(j)}\rangle \langle b^{(j)} \vert  U^{(j)}_{i}],
\end{align}
in the same spirit as the definition of $\wt{\mc{M}}$, with,
\begin{align} \label{eq:local_estim_channel}
    \wt{\mc{M}}_{\mathrm{local}}(\rho) = 3\rho - \openone.
\end{align}

As the effect of $\mc{M}_{\mathrm{local}}$ is to
contract the single-qubit Bloch sphere uniformly by a factor of 3,
the inverse channel is given by dilating the Bloch sphere by the same factor \cite{huang2020predicting}.
Our scheme biases this channel by effectively dilating the Bloch sphere by a smaller factor tuned via a bias parameter $\varepsilon$.
\begin{statement}[biased shadow estimators]\label{stat:main_result}
Given a bias parameter $\varepsilon$ 
we modify the conventional shadow estimator in \cref{eq:local_estim_channel}
and define the biased local estimator of
Pauli shadows $\wt{\mc{M}}^{(\varepsilon)}_{\mathrm{local}}$ 
as
\begin{align}
	\wt{\mc{M}}^{(\varepsilon)}_{\mathrm{local}}(\rho) = 3\sqrt{1 - \varepsilon} \rho + \frac{1}{2} \left(1 - 3 \sqrt{1 - \varepsilon}\right) \mathds{1} \mathrm{.}
\end{align}

This channel dilates the Bloch  sphere by a factor $3\sqrt{1 - \varepsilon}$
and, indeed, for $\varepsilon = 0$ we recover the unbiased channel of conventional classical shadows.
\end{statement}
While the above channel does not converge to the true state in the infinite-sample limit for fixed (independent of $N_s$) $\varepsilon \neq 0$, it allows us to control shot noise (statistical uncertainty) in the finite-sample regime as the variance is decreased by $(1-\varepsilon)$. 
As we argue, by choosing the optimal $\varepsilon$ as a function of $N_s$,
we obtain a consistent estimator, which converges to the unbiased estimate in the infinite-sample limit.

We now show that this trade-off is on average worthwhile through
defining the expected loss as a measure of the \emph{average} performance of the biased scheme as
\begin{equation}
	\mc{L}_{\mc{U}}(\rho, \varepsilon) = \mathds{E}_{\mc{U}, \mathbf{b}} \{\tr[(\rho - \hat{\rho})^2]\}.
\end{equation}
In \cref{fig:densityplot}, we consider applying the Pauli shadows
approach to estimating local properties of a quantum system and consider
the reduced density matrix of a single-qubit.
We fix a number of shots (samples) $N_s = 100$, a biasing parameter of $\varepsilon = 0.1$, and the true state $\rho$ on the $Z$-axis of the Bloch sphere. 
We calculate the change in loss for a set of averaged classical-shadow estimates of $\rho$ under biasing by $\varepsilon$.
The exterior of the dashed circle corresponds to estimates where the corresponding biased estimator is actually more accurate in the particular finite-sample regime than the original, unbiased one. 
We now concretely state our analytical result that quantifies the expected loss for any
particular local density matrix.
\begin{figure}
	\centering
	\includegraphics[width=\linewidth]{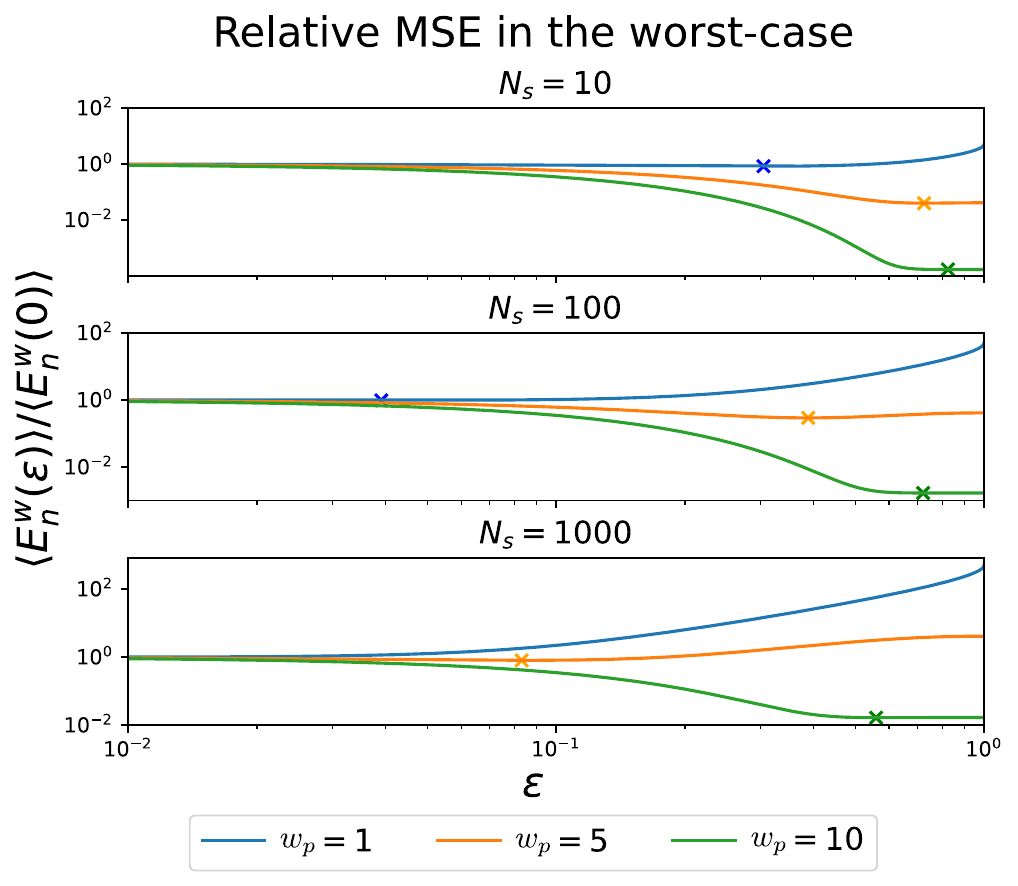}
	\caption{
		Mean squared error $\langle E_{n}^{\text{w}}(\varepsilon) \rangle $ of
		the biased scheme relative to the unbiased one when the aim is
		to estimate Pauli expected values of a weight $w_p$ observable on an $n$-qubit state $\tr(O \rho)$.
		Analytical expression plotted for the worst case scenario ($O$ commutes with $\rho$)
		for an increasing number of shots, weight of the Pauli string ($n = w_p$) and bias parameters $\varepsilon$.
		There is an optimal $\varepsilon$ (crosses at the minimum of the curves) for which the relative error is minimised. When $w = 1$ and $N_s = 1000$ the minimum is attained outside the plotted region.
		}
	\label{fig:mse_worst_case}
\end{figure}

\begin{statement}[single-shot average-case loss]\label{statement:loss}
For a single-qubit local density matrix $\rho = \tfrac{1}{2}[\mathds{1} + (\boldsymbol{r} \cdot \boldsymbol{\sigma})]$,
where $\mc{U}$ forms a 2-design, the expected loss for a single sample is
\begin{align}
	\mc{L}_{\mc{U}}(\rho, \varepsilon) = \frac{1}{2} \left[\lVert \boldsymbol{r} \rVert^2 + 9(1 - \varepsilon)\right] - \sqrt{1 - \varepsilon} \lVert \boldsymbol{r} \rVert^2.
\end{align}
See \cref{app:avg_loss} for the derivation.
This is a quadratic expression in $\sqrt{1 - \varepsilon}$ and thus
attains a minimum value at $\sqrt{1 - \varepsilon_{\mathrm{min}}} = \frac{1}{9}\lVert \boldsymbol{r} \rVert^2$ as
\begin{align}
	\mc{L}_{\mc{U}}(\rho, \varepsilon_{\mathrm{min}}) = \frac{1}{18} \lVert \boldsymbol{r} \rVert^2 \left(9 -  \lVert  \boldsymbol{r} \rVert^2\right)
\end{align}

\end{statement}
Thus, even if $\rho$ is pure ($\lVert \boldsymbol{r} \rVert = 1$),
it is worth introducing a strong bias for a single shot estimator as $\varepsilon_{\mathrm{min}} = \frac{80}{81}$ 
because it reduces the expected loss to $4/9$ from the expected loss $4$ of the unbiased scheme.
However, in practice the qubit is usually part of an entangled (computational) state ($\lVert \boldsymbol{r} \rVert < 1$)
	which guarantees even more significant gains.
In the following sections we explain, however, that the gain is less pronounced as we increase the number of shots.

\subsection{Analysing worst- and best-case gain}
While in \cref{statement:loss} we focused on the average performance 
of biased shadow tomography, we
now analytically predict the performance in extremal scenarios. 
In particular, we consider the practically pivotal task of predicting expected values of local Pauli observables from the snapshots 
as $\tr(O \hat{\rho})$ and analyse the worst- and best-case scenarios.
We note that we focus on the mean estimator, which is then of course the primary
component of the median-of-means estimator~\cite{huang2020predicting}.

As we will conclude below \cref{eq:biasing_gain_shadows},
our biased estimator yields the least gain in the worst-case scenario that the quantum state $\rho$
is the eigenstate of the Pauli observable via $\tr(O \rho) = +1$ (where $O$ is a Pauli string of weight $w_p$ up to $\pm$ sign). 
The reason why we still gain even in this worst-case scenario is the following:
 at each shot our estimator $\tr(O \hat{\rho})$ yields the outcome 
either $+ 1$ (when we measure in a compatible Pauli basis) or $0$ (when we measure in an incompatible basis). 
This indeed yields a binomial distribution $B(N_s,p)$ with number of samples $N_s$ and 
probability of compatible measurements $p=3^{-w}$.
As we illustrate in the Appendix (\cref{fig:binomial}), the binomial distribution is not symmetric around the mean
but has a tail, and the mean squared error on the right-hand side is higher than the mean squared error on the
left-hand side -- it is thus always worth biasing the estimator by effectively rescaling the estimate.
However, as we increase $N_s$, the binomial distribution quickly
tends to a symmetric normal distribution and thus rescaling will not yield an advantage.

In \cref{app:single_qubit_analysis}
we analytically derive the mean error
$ \langle E_{n}^{\text{w}}(\varepsilon) \rangle$ in the expected value measurement and 
plot these relative to the unbiased case in \cref{fig:mse_worst_case}.
Indeed, this confirms that 
(a) The biased scheme is always advantageous as there is always an optimal
$\varepsilon$ (the minimum of the curves indicated using crosses) for which the relative mean error is smaller than $1$.
(b) The advantage of the biased scheme grows exponentially as we increase
the weight of the Pauli string simply because the
contraction of the Bloch sphere is exponential via the factor $3^{w_p}$
for locality $w_p$ while the tail of the distribution of estimates (via the aforementioned binomial distribution)
gets exponentially long, i.e., compare blue, orange and green lines.
(c) The advantage of the biased scheme diminishes as we increase the numbers of shots.

In contrast, the best case scenario is attained when the quantum state is an eigenstate
of an operator that anticommutes with the observable which gives us $\Tr(O\rho)=0$. It is then clear that biasing, which is equivalent to shrinking the expectation value is always advantageous 
as it forces the estimate to be closer to $0$.
In \cref{app:single_qubit_analysis}, we give an expression for the mean error $\langle E_{n}^{\text{b}}(\varepsilon)\rangle$ which we numerically compute and plot in \cref{fig:mse_best_case}
confirming that indeed
increasing the bias parameter $\varepsilon$ monotonically decreases the expected error.

\subsection{\label{sec:var_qm_expec} Relation to variances and quantum mechanical expected values}
In the previous section we analysed the instances when the expected values attain the extremal values
$\tr(O \rho) = \pm 1$ and $\tr(O \rho) = 0$, and  we now prove that indeed these are the worst- and best-case scenarios, respectively.
For this reason we consider an arbitrary sample mean estimator $\overline{R}$ that one obtains
from averaging over $N_s$ samples of a single-shot estimator $\hat{R}$,
and compare $\overline{R}$ to the corresponding biased estimator $(1-\alpha)\overline{R}$
that one obtains through rescaling with the factor $(1-\alpha)$.
In order to derive the optimal biasing point, i.e., the minimum of the curves in \cref{fig:mse_worst_case} (crosses),
we first consider the Mean Squared Error (MSE) of the unbiased estimator $\overline{R}$
as $\mse{\overline{R}} = \var{\hat{R}}/N_s$. 
We can define and calculate the Signal to Noise Ratio (SNR) of $\overline{R}$ as
\begin{equation}\label{eq:SNR}
		\beta := \frac{\expect{\overline{R}}^2}{\mse{\overline{R}}} =  \frac{\expect{\hat{R}}^2}{\var{\hat{R}}/N_s}.
\end{equation}
The MSE for the biased mean estimator $(1-\alpha)\overline{R}$ is then
\begin{equation}
	     \mse{(1-\alpha) \overline{R}} 
	 =  \underbrace{\alpha ^2 \expect{\hat{R}}^2}_{\text{bias}} +  \underbrace{(1-\alpha)^2\var{\hat{R}}/N_s}_{\text{variance}},
\end{equation}
which is minimised at the optimal biasing point $\alpha^*~=~(1+\beta)^{-1}$ as we derive in \cref{app:stats}. 
Through \cref{eq:SNR}, we find that the optimal bias parameter approaches $\alpha^* \rightarrow 0$ as we increase the number of samples $N_s \rightarrow \infty$.  As such, our (optimally)
biased estimator is actually a consistent estimator, i.e., it asymptotically approaches an unbiased estimator in the infinite-sample limit~\cite{scott1981asymptotic}.

We can make the following statement at the optimal biasing point.

\begin{statement}[biasing an estimator through rescaling]
	The SNR of the optimally-biased estimator is
	$\beta_{\mathrm{biased}} = 1+\beta$ which always guarantees an improved
	SNR over the unbiased estimator $\beta$. The relative SNR gain through biasing is given as
	\begin{equation}
		\frac{ \beta_{\mathrm{biased}}  } {\beta} = 1 + \beta^{-1}
    \label{eq:biasing_gain}
	\end{equation}
 \label{stat:optimal_bias}
\end{statement}

As further shown in \cref{app:stats}, in the case of estimating Pauli expected values for Pauli shadows, the mean and variance of the single-shot estimator are given as $\expect{\hat{R}} = \tr[O \rho]$ and $\var{\hat{R}} =3^{w}  -  \tr[O \rho]^2$. Hence, the SNR of the unbiased estimator is given as  $\beta = N_s [ 3^w\tr[O \rho]^{-2}  - 1 ]^{-1}$ and the factor of SNR gain by biasing is given by:
\begin{equation}\label{eq:biasing_gain_shadows}
		\frac{ \beta_{\mathrm{biased}}  } {\beta} = 1 + (3^w\tr[O \rho]^{-2}  - 1)/N_s
	\end{equation}
The maximal and minimal gains are then obtained at $\tr[O \rho] = 0$ and  $\tr[O \rho] = +1$ respectively, which proves our previous observations on worst- and best-case scenarios.

Let us note that, while the above SNR allows us to rigorously prove that biasing is in principle always advantageous,
the mean-squared error analysed in the previous section remains
the more practical measure and we will use it thereafter.
It is also worth noting that \cref{stat:optimal_bias} requires us to know the expectation
value $\expect{\hat{R}}$ exactly---which may not be possible in practice---in order to predict the optimal biasing point.
Nevertheless, we demonstrate in the following section
that biasing is still worthwhile even if we can only have approximate knowledge of the optimal bias.

\section{Practical demonstration}
We consider a potential practical application whereby
one aims to obtain the ground-state energy of a Hamiltonian $\mathcal{H}$ composed of
only low-weight (local) Pauli observables.
In an experiment one first prepares the ground state and 
collects a set of Pauli shadows from which the ground-state energy can be predicted in post-processing.
The significant advantage of classical shadows is that they allow
us to estimate further Pauli strings beyond the Hamiltonian terms,
without repeating the experiment.

For example, one can consider
perturbative corrections to the Hamiltonian in the form of high-weight Pauli strings $P$
and predict the expected value of the sum $\mathcal{H} + P$, such as when 
computing a first-order correction to the energy in perturbation theory.
However, the variance in our example is increased exponentially due to the high weight of $P$;
this potentially renders a direct estimation impractical as
the increased variance may bring the SNR down below $1$, as we detail in \cref{app:application}.
As we demonstrate, biasing then allows us to significantly improve upon this potentially low SNR.

As a concrete example we consider a spin-ring Hamiltonian as
\begin{equation}\label{eq:spin_ring_hamiltonian}
    \mathcal{H} = \sum_{k \in \text{ring}(N)} \omega_k Z_k^{z} + J \boldsymbol{\sigma}_k\cdot\boldsymbol{\sigma}_{k+1},
\end{equation}
with coupling $J = 0.3$, on-site interaction strengths uniformly randomly
generated in the range $-1 \leq \omega_k \leq 1 $ and $\boldsymbol{\sigma}_k = (X_k, Y_k, Z_k)^T$
is a vector of single-qubit Pauli matrices.
We prepare the ground state $\rho$ of a $12$-qubit Hamiltonian
through a variational Hamiltonian ansatz of $l = 5$ layers. We generate a collection of shadows $S(N_s, \rho)$ with $N_s = 10^6$ and estimate the expected value $\tr[(\mathcal{H}+P)\rho]$
with respect to corrections $P$ as $w=8$ Pauli observables.

\begin{figure}
    \centering
    \includegraphics[width=\linewidth]{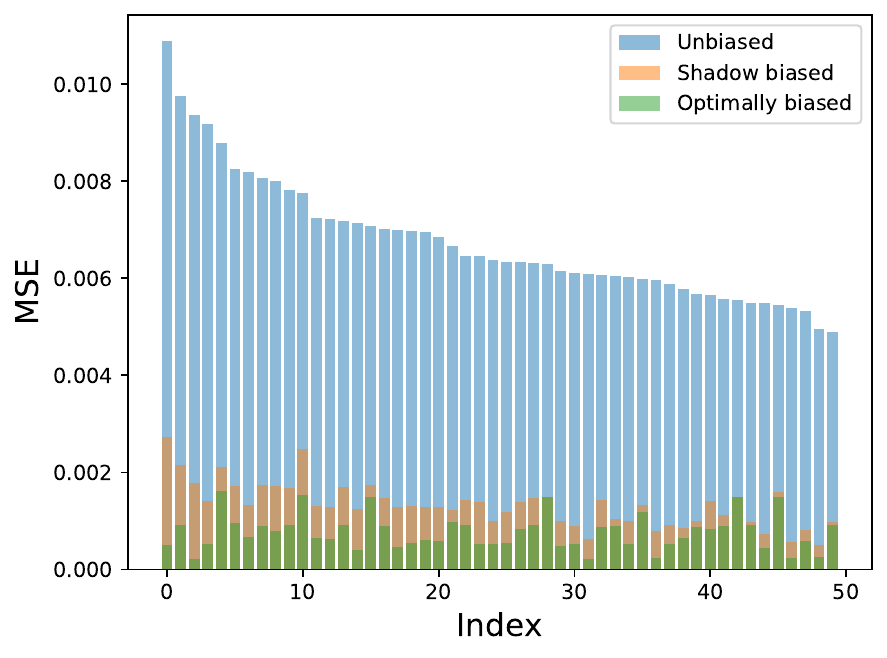}
    \caption{Illustration of the impact of different biasing strategies. In blue, we plot the MSE of fifty 8-local observables (whose SNR is smaller than $1$) sorted by how large the error is. While analytically choosing the optimal bias according to \cref{stat:optimal_bias} drastically reduces the error (green), this assumes previous knowledge of the exact expectation values, which is not available in practice. Through estimating the optimal bias from experimental data (orange), 
    	one still obtains a significant error reduction.}
    \label{fig:mse_opt_shadow_bias}
\end{figure}

In \cref{fig:mse_opt_shadow_bias}, we plot the mean-squared error
obtained by averaging over $10^5$ repetitions and consider different $8$-local
observables with and without biasing (blue). We consider two different
biasing strategies. First, we analytically choose the optimal bias parameter $\alpha^*$
according to \cref{stat:optimal_bias} (green) assuming direct access to the exact expected values
(which one does not have access to in practice).
Second, we use the experimentally estimated expected values using shadows to estimate the 
 optimal biasing point in \cref{stat:optimal_bias}.
While the latter deviates from the exact $\alpha^*$ due to shot noise,
\cref{fig:mse_opt_shadow_bias} (orange) clearly demonstrates that the MSE is still significantly reduced compared
to the unbiased scheme.
This confirms robustness against errors in our ability to determine the optimal bias parameter,
i.e., the approach can demonstrably be used even when the optimal biasing point can only be estimated from
the experimental data.

\section{Discussion and Conclusion}
In this work, we explore the tradeoff between intentionally introducing a bias into 
classical shadows estimators 
which in return allows us to reduce statistical uncertainties due to finite samples -- 
ultimately enabling us to achieve the same precision but using fewer samples.
The implementation of the approach is straightforward, and is performed completely
in post-processing, as it is effectively just a rescaling of the conventional shadow estimator.

We obtain rigorous analytical guarantees that 
biasing is, in principle, always worthwhile given an optimal bias parameter is known:
First, optimal biasing improves the relative loss of shadow tomography on average.
Second, optimal biasing improves expected value measurements even in the worst-case scenario
and may provide significant gains in other scenarios.
Third, optimal biasing is guaranteed to increase the signal-to-noise of any statistical mean estimator.
Although the optimal bias parameter is not known a priori,
we demonstrate in a practically motivated numerical experiment that an approximation directly determined form experimental
data is sufficient. 

The gain using our biased estimator is more pronounced
for a small number of samples or  when Pauli strings of high weight are predicted.
As such, our approach is thus particularly well suited for practical tasks 
where only a relatively low number of shots is available, e.g.,
estimating gradients \cite{schuld2019evaluating,andesc,wierichs2022general,natgrad} or covariances
\cite{PhysRevX.12.041022} when training variational circuits or when estimating time-dependent properties~\cite{chan2022algorithmic}. Another particularly interesting application area is quantum error mitigation~\cite{cai2023quantum} whereby
prior works aimed at recovering an unbiased estimator from noisy measurements.
Combinations of QEM and classical shadows have similarly been considered~\cite{jnane2023quantum,zhao2023group}.

The insights and theoretical results provided in this work may prove invaluable
in developing further, more advanced biased estimators.
Given classical information from quantum systems can only be extracted through statistical
estimators, this work is an important step in the crucial task of developing applications of quantum computers
that have minimal sample requirements.

\section*{Acknowledgments}
The numerical modelling involved in this study made
use of the Quantum Exact Simulation Toolkit (QuEST)~\cite{quest}, and the recent development
QuESTlink~\cite{QuESTlink} which permits the user to use Mathematica as the
integrated front end, and pyQuEST~\cite{pyquest} which allows access to QuEST from Python.
We are grateful to those who have contributed
to all of these valuable tools. 
The authors would like to acknowledge the use of the University of Oxford Advanced Research Computing (ARC)
facility~\cite{oxford_arc} in carrying out this work
and specifically the facilities made available from the EPSRC QCS Hub grant (agreement No. EP/T001062/1).
The authors also acknowledge funding from the
EPSRC projects Robust and Reliable Quantum Computing (RoaRQ, EP/W032635/1)
and Software Enabling Early Quantum Advantage (SEEQA, EP/Y004655/1). 
B.K. thanks UKRI for the Future Leaders Fellowship
project titled Theory to Enable Practical Quantum Advantage (MR/Y015843/1).
B.K. thanks the University of Oxford for
a Glasstone Research Fellowship and Lady Margaret Hall, Oxford for a Research Fellowship. Z.C. is supported by the Junior Research Fellowship from St John’s College, Oxford. A.C. acknowledges support from EU H2020-FETFLAG-03-2018 under grant agreement no. 820495 (AQTION).

\clearpage
\onecolumngrid

\appendix

\section{Comparison to previous literature}

The idea of introducing a bias in the context of classical shadows has been explored in several previous works, and here we compare our results to some notable examples.

In Ref.~\cite{hadfield2022obtaining}, the authors consider biasing the local measurement bases at the measurement stage of preparing classical shadows (i.e. choosing measurements from a biased distribution $\mc{U}$ over local basis rotations $U_i$).
These authors then choose the biasing distribution to minimize the variance and report an unbiased estimator for the energy of a molecular Hamiltonian, expressed as a sum of Pauli strings. 
In Ref.~\cite{gresch2023guaranteed}, the authors introduce a biased estimator for the energy of a Pauli Hamiltonian by truncating the set of Pauli estimators included in the sum of the energy estimator.
This bias is chosen based on the number of compatible measurement settings in a given measurement scheme, and is guaranteed to achieve the optimal tradeoff between the statistical error and the systematic error introduced by biasing.
In both Refs.~\cite{hadfield2022obtaining} and \cite{gresch2023guaranteed}, the authors are considering estimating a single observable given by a sum of Pauli operators, as opposed to estimating multiple independent observables simultaneously.

In Ref.~\cite{zhu2023connection}, the authors show that least-squares and regularized least-squares estimators can be viewed as instances of classical shadows that are generally biased.
The authors note that an essential conceptual difference between the least-squares estimators and conventional classical shadows is that classical shadows make use of a hypothetical distribution from which unitary rotations are drawn.
Our approach can be seen as incorporating this source of randomness together with bias.

\section{\label{app:avg_loss} Average case loss}
In this section, we show the following result for the expected loss of a single qubit
\begin{align}
    \mathds{E}_{\mc{U}, \mathrm{b}} \{\tr[(\rho - \hat{\rho})^2]\} = \frac{1}{2} [\lVert \boldsymbol{r} \rVert^2 + 9(1 - \varepsilon)] - \sqrt{1 - \varepsilon} \lVert \boldsymbol{r} \rVert^2 \mathrm{,}
\end{align}
where 
\begin{align}
    \hat{\rho} = \mc{M}^{(\varepsilon)} \left[U \ketbra{\mathrm{b}}{\mathrm{b}} U^{\dagger} \right]\mathrm{,} \mbox{\hspace{10mm}}
    \rho = \frac{1}{2}(I + \boldsymbol{r} \cdot \boldsymbol{\sigma}) \mathrm{,}
\end{align}
and $\mc{U}$ is a Haar 2-design.
We have
\begin{align}
\mathds{E}_{\mc{U}, \mathrm{b}} \{\tr[(\rho - \hat{\rho})^2]\}  &= \sum_{\mathrm{b} \in \mathds{Z}_2} \int_{\mc{U}}
 dU \ \bra{\mathrm{b}} U^{\dagger} \rho U \ket{\mathrm{b}} \tr\{[\rho - \mc{M}^{(\varepsilon)}(U \ketbra{\mathrm{b}}{\mathrm{b}} U^{\dagger})]^2\}
\end{align}
Since $\mc{M}^{(\varepsilon)}$ only rescales the Bloch vector, the identity parts of $\rho$ and $\hat{\rho}$ cancel, and we have
\begin{align}
\mathds{E}_{\mc{U}, \mathrm{b}} \{\tr[(\rho - \hat{\rho})^2]\}  &= \frac{1}{4} \sum_{\mathrm{b} \in \mathds{Z}_2} \int_{\mc{U}} dU \ \bra{\mathrm{b}} U^{\dagger} \rho U \ket{\mathrm{b}} \tr\{[\left(\boldsymbol{r}\cdot \boldsymbol{\sigma} \right) - 3 (-1)^{\mathrm{b}} \sqrt{1 - \varepsilon} \left(U Z U^{\dagger}\right)]^2\} \\
\mathds{E}_{\mc{U}, \mathrm{b}} \{\tr[(\rho - \hat{\rho})^2]\}   &= \frac{1}{4} \sum_{\mathrm{b} \in \mathds{Z}_2} \int_{\mc{U}} dU \ \bra{\mathrm{b}} U^{\dagger} \rho U \ket{\mathrm{b}} \tr\{[\lVert \boldsymbol{r} \rVert^2 + 9(1 - \varepsilon)]I - 6 (-1)^{\mathrm{b}} \sqrt{1 - \varepsilon} \left(U Z U^{\dagger} \right) \left(\boldsymbol{r}\cdot \boldsymbol{\sigma} \right) \}
\end{align}
Inside of the trace, we have a term which is proportional to the identity, and a term which is proportional to the operator $(U Z U^{\dagger})(\boldsymbol{r} \cdot \boldsymbol{\sigma})$.
Calculating each one at a time, we have
\begin{align}
(\mathrm{identity \ term}) &= \frac{1}{4} \sum_{\mathrm{b} \in \mathds{Z}_2} \int_{\mathcal{U}} dU \ \bra{\mathrm{b}} U^{\dagger} \rho U \ket{\mathrm{b}} \tr\{[\lVert \boldsymbol{r} \rVert^2 + 9(1 - \varepsilon)]I \} \\
&= \frac{1}{2}[\lVert \boldsymbol{r} \rVert^2 + 9(1 - \varepsilon)]
\end{align}
and
\begin{align}
(\mathrm{non-identity \ term}) &= -\frac{3}{2} \sqrt{1 - \varepsilon} \sum_{\mathrm{b} \in \mathds{Z}_2} (-1)^{\mathrm{b}} \int_{\mathcal{U}} dU \ \bra{\mathrm{b}} U^{\dagger} \rho U \ket{\mathrm{b}} \tr\{\left(U Z U^{\dagger} \right)\left(\boldsymbol{r}\cdot \boldsymbol{\sigma} \right) \}\\
&= -\frac{3}{2} \sqrt{1 - \varepsilon} \sum_{\mathrm{b} \in \mathds{Z}_2} (-1)^{\mathrm{b}} \tr\left\{\mathcal{E}^{(2)} \left(\ketbra{\mathrm{b}}{\mathrm{b}} \otimes Z \right) \left[\rho \otimes \left(\boldsymbol{r} \cdot \boldsymbol{\sigma} \right)\right]\right\} \\
&= -\frac{3}{2} \sqrt{1 - \varepsilon} \tr\left\{\mathcal{E}^{(2)} \left(Z \otimes Z \right) \left[\rho \otimes \left(\boldsymbol{r} \cdot \boldsymbol{\sigma} \right)\right]\right\} \\
&= -\frac{1}{2} \sqrt{1 - \varepsilon} \tr\left\{ \left(2 S - I \right) \left[\rho \otimes \left(\boldsymbol{r} \cdot \boldsymbol{\sigma} \right)\right]\right\} \\
(\mathrm{non-identity \ term})  &= -\sqrt{1 - \varepsilon} \lVert \boldsymbol{r} \rVert^2
\end{align}
where $\mathcal{E}^{(2)}(A) = \int_{\mathcal{U}} dU U^{\otimes 2} A (U^{\otimes 2})^{\dagger}$, and $S$ is the 2-qubit SWAP operation. 
Therefore
\begin{align}
\mathds{E}_{\mc{U}, \mathrm{b}} \{\tr[(\rho - \hat{\rho})^2]\} = \frac{1}{2} [\lVert \boldsymbol{r} \rVert^2 + 9(1 - \varepsilon)] - \sqrt{1 - \varepsilon} \lVert \boldsymbol{r} \rVert^2
\end{align}
This is a quadratic expression in $\sqrt{1 - \varepsilon}$, so it attains a minimum value at
\begin{align}
\sqrt{1 - \varepsilon_{\mathrm{min}}} = \frac{1}{9}\lVert \boldsymbol{r} \rVert^2 \mbox{\hspace{5mm}} \mathrm{with} \mbox{\hspace{5mm}} \mathds{E}_{\mc{U}, \mathrm{b}} \{\tr[(\rho - \widetilde{\rho})^2]\}|_{\varepsilon_{\mathrm{min}}} = \frac{1}{2} \lVert \boldsymbol{r} \rVert^2 - \frac{1}{18}\lVert \boldsymbol{r} \rVert^4
\end{align}
Therefore, we can in-fact minimize the expected loss and decrease the classical-shadow sampling variance by introducing a biased inverse channel.

\section{Analysing worst and best-case scenarios\label{app:single_qubit_analysis}}
\subsection{Single qubit error\label{app:single_qubit_error}}

Figure \ref{fig:densityplot} depicts an emperical distribution of Bloch-vector estimates for a single qubit, where each pixel is taken as a Gaussian kernel density estimate of the average over $N_s = 100$ samples. 
Unitaries are drawn from $\mc{U} = \mc{C}_1$ and projected onto the $\avg{X}$, $\avg{Z}$ plane.
Each of the estimates in this plot is unbiased, and we consider how biasing by $\varepsilon = 0.1$ as described in the previous section would hypothetically affect every estimate.
The exterior of the dashed circle in the figure is the region where the biased estimate has a lower distance to the true mean value than the unbiased one.
As we increase $\varepsilon$, the radius of the circle increases such that the origin and the unbiased mean lie on a diameter of the circle.
When we include the $\avg{Y}$ value, this circle becomes a sphere of revolution about the common diameter defined by the origin and the unbiased mean.

We obtain an expected advantage from biasing our estimate if the increase in loss due to the contribution from estimates inside of the dashed circle is compensated by the corresponding decrease in loss from estimates outside the circle.
Intuitively, this is possible due to the geometry of the sphere and the fact that biasing pulls the estimates to the center.
The region outside of the dashed circle includes a region where the error in all three Bloch vector components is decreased by biasing, whereas for every estimate inside the dashed circle, the error in at most one Bloch vector component is increased.

\subsubsection{Worst case}

The worst case scenario happens when the studied quantum state $\rho$ is an eigenstate of the Pauli observable of interest $O$. For ease of presentation, let us focus first on the single-qubit case. We generalise the following results to $n$-qubit states in the next subsection. Without loss of generality, let $\rho = \ketbra{+}{+}$ and $O = X$ which thus satisfies $\tr(O\rho) = 1$. For a given collection of shadows $S(\rho, N_s) = \{\hat{\rho}_1, \dots, \hat{\rho}_{N_s} \}$, our estimate for the expectation value is given by the mean of each snapshot's expectation value, 
\begin{align}
    \frac{1}{N_s}\sum_{i = 1}^{N_s} \tr(X\hat{\rho}_i).
\end{align}
Each $\hat{\rho}_i$ has been obtained by measuring in a randomly chosen single-qubit
Pauli basis, and as a result $\tr(X\hat{\rho}_i)$ yields
either $+ 1$ (when we measure in the Pauli $X$ basis) or $0$ (when we measure in the $Y$ or $Z$ bases).
Here we aim to calculate the mean error made in computing the expectation value $\langle E_{1}^{\text{w}}(\varepsilon) \rangle$ when using a bias parameter $\varepsilon$.
Suppose that $k$ snapshots resulted in the outcome $+1$ while
$N_s - k $ resulted in the outcome $0$, 
we define the error in the mean value as,
\begin{align}
    E_{1}^{\text{w}}(\varepsilon) &\coloneqq \left|\frac{1}{N_s}\sum_{i = 1}^{N_s} \tr(X\hat{\rho}_i)-1\right|^2 = \left|\frac{3\sqrt{1-\varepsilon}}{N_s}k-1\right|^2.    
\end{align}
The average error is then given by,
\begin{align}
    \langle E_{1}^{\text{w}}(\varepsilon)\rangle = \sum_{k=0}^{N_s}P\left(\sum_{i = 1}^{N_s}\tr(X\hat{\rho}_i) = k\right)\left|\frac{3\sqrt{1-\varepsilon}}{N_s}k-1\right|^2,
\end{align}
where $\sum_{i = 1}^{N_s}\tr(X\hat{\rho}_i)$ is simply a random variable following a binomial distribution
$B(N_s, p)$ with $p=\tfrac{1}{3}$. We can then obtain an analytic expression for
$P\left(\sum_{i = 1}^{N_s}\tr(X\hat{\rho}_i) = k\right)$ leading to,
\begin{align}
    \langle E_{1}^{\text{w}}(\varepsilon)  \rangle= \sum_{k=0}^{N_s}{N_s\choose{k}}p^k(1-p)^{(N_s-k)}\left|\frac{3\sqrt{1-\varepsilon}}{N_s}k-1\right|^2.
\end{align}

\begin{figure}
    \centering
    \includegraphics[width = 0.6\linewidth]{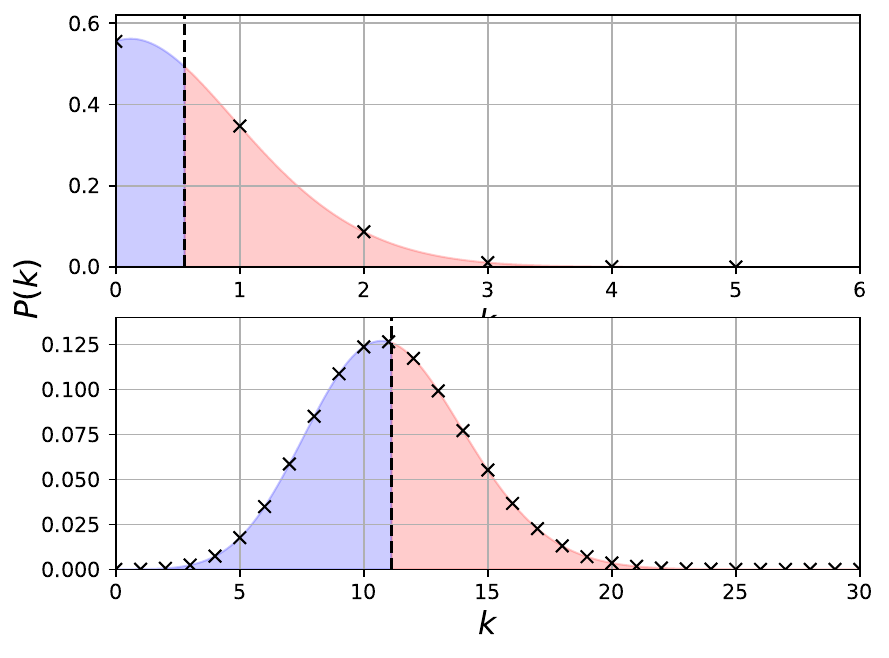}
    \caption{Binomial distribution for $p= 1/9$, $N_s = 5$ (top) and $N_s = 100$ (bottom). For a small $N_s$, it is more likely to obtain an outcome larger than the mean (black dashed line) which leads to a high mean-squared error which can be reduced by biasing the estimator. When $N_s$ gets larger, biasing doesn't improve our estimates as the distribution becomes symmetric around the mean.}
    \label{fig:binomial}
\end{figure}

\subsubsection{Best case}
\begin{figure}
    \centering
    \includegraphics[width=0.5\textwidth]{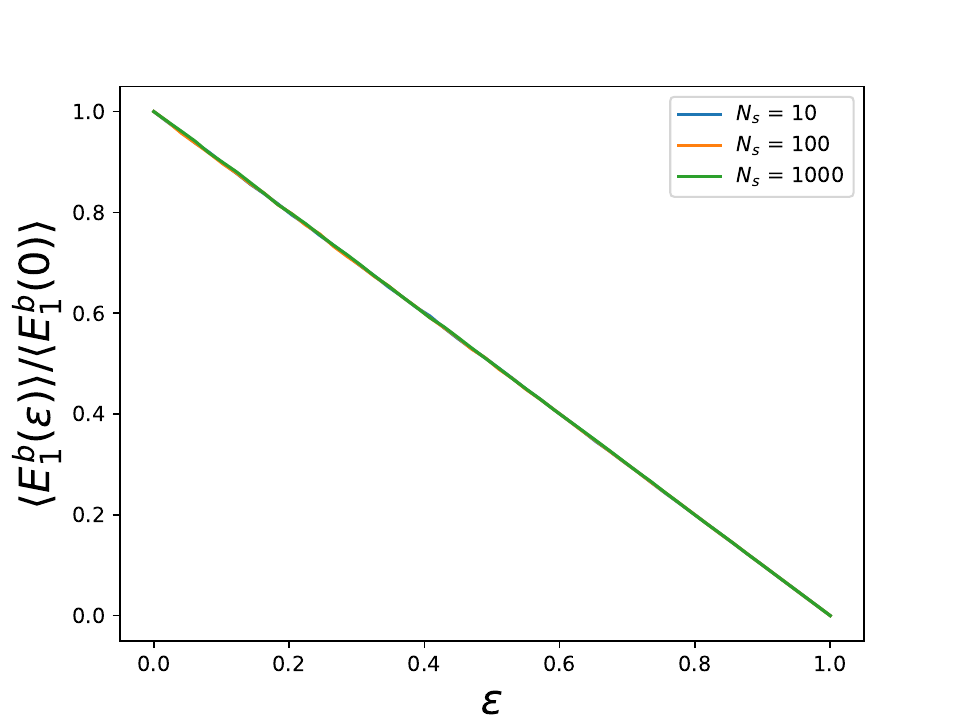}
    \caption{Numerical evaluation of the mean squared error $\langle E_{1}^{\text{b}}(\varepsilon) \rangle $ of
		the biased scheme relative to the unbiased one when the aim is
		to estimate Pauli expected values $\tr(O \rho)$ in the best-case scenario
		for an increasing number of shots $N_s$ and bias parameters $\varepsilon$.
		As $\tr(O \rho) = 0$ and biasing corresponds to shrinking the expectation value, we always reduce the relative error.}
    \label{fig:mse_best_case}
\end{figure}

The best case scenario is obtained when the state is an eigenstate of an operator that
anticommutes with our observable, such as $\rho = \ketbra{+}{+}$ and
 $O = Z$ which gives us $\tr(O\rho) = 0$.
For each sample our estimator yields either $+1$ or $-1$ with equal probability
(compatible measurement in the Pauli $Z$ basis) or yields $0$ (incompatible measurement via Pauli $X$ and $Y$ bases).
Again, we can compute the mean error over the distribution of $k$ as
\begin{align}
    \langle E_{1}^{\text{b}}(\varepsilon) \rangle = \sum_{k=-N_s}^{N_s}P\left(\sum_{i = 1}^{N_s}\tr(Z\hat{\rho}_i) = k\right)\left|\frac{3\sqrt{1-\varepsilon}}{N_s}k\right|^2.
    \label{eq:best_case_mse}
\end{align}
With $ P(\tr(Z\hat{\rho}_i) = 0) = \tfrac{2}{3}$ and $P(\tr(Z\hat{\rho}_i) = 1) = P(\tr(Z\hat{\rho}_i) = -1) = \tfrac{1}{6}$. Finding an analytic formula for $P\left(\sum_{i = 1}^{N_s}\tr(Z\hat{\rho}_i) = k\right)$ is quite challenging, but one can easily compute it numerically. In \cref{fig:mse_best_case}, we compute the relative mean squared error between the biased and unbiased cases by sampling \cref{eq:best_case_mse} $N_{reps} = 10^6$ times for different values of $N_s$. The plot confirms our intuition that increasing the bias monotonically decreases the error.

\subsection{Multiqubit error}
The generalisation to the multiqubit case is straightforward as both the initial state and observable can be written as tensor products. 
Let $\rho = \ketbra{+}{+}^{\otimes n}$ and $O = X^{\otimes n}$, with $n$ the number of qubits.
For a given collection of shadows $S(\rho, N_s) = \{\hat{\rho}_1, \dots, \hat{\rho}_{N_s} \}$ such that $\forall i, \hat
{\rho}_i = \otimes_{j=1}^{n}\hat{\rho}_{i,j}$ the error in the worst-case now reads, 
\begin{align}
    \langle E_n^{\text{w}}(\varepsilon)\rangle = \sum_{k=0}^{N_s}P\left(\sum_{i = 1}^{N_s}\left(\prod_{j=1}^{n}\tr(X\hat{\rho}_{i,j})\right) = k\right)\left|\frac{3^n(1-\varepsilon)^{n/2}}{N_s}k-1\right|^2.
\end{align}

Here $\sum_{i = 1}^{N_s}\left(\prod_{j=1}^{n}\tr(X\hat{\rho}_{i,j})\right) \sim B(N_s, p_{n})$ with $p_{n} = \tfrac{1}{3^n}$. Hence, 
\begin{align}
    \langle E_n^{\text{w}}(\varepsilon)\rangle = \sum_{k=0}^{N_s}{N_s\choose{k}}p_{n}^k(1-p_n)^{(N_s-k)}\left|\frac{3^n(1-\varepsilon)^{n/2}}{N_s}k-1\right|^2.
\end{align}
Plotting $\langle E_n^{\text{w}}(\varepsilon)\rangle$ in \cref{fig:mse_worst_case} for different scenarios, we find that it is always worth biasing our estimator. 

In the best case, with $\rho = \ketbra{+}{+}^{\otimes n}$ and $O = Z^{\otimes n}$ we find,

\begin{align}
    \langle E_n^{\text{b}}(\varepsilon)\rangle = \sum_{k=-N_s}^{N_s}P\left(\sum_{i = 1}^{N_s}\left(\prod_{j=1}^{n}\tr(Z\hat{\rho}_{i,j})\right) = k\right)\left|\frac{3^n(1-\varepsilon)^{n/2}}{N_s}k\right|^2,
\end{align}
Here again, one will need to numerically compute the error.

\section{Statistical Analysis for Biased Shadow\label{app:stats}}
\subsection{Biasing an estimator by rescaling}
\label{sec:varbiastrade}
Suppose the target parameter that we want to estimate can be obtained using some \emph{unbiased} estimator $\hat{R}$ with the corresponding sample mean estimator after $N_s$ shots denoted as $\overline{R}$. In this way, the mean square error (MSE) of the this unbiased estimator after $N_s$ shots is simply
\begin{align}\label{eqn:unbias_mse}
    \mse{\overline{R}} = \var{\overline{R}} =  \var{\hat{R}}/N_s.
\end{align}
The error of the unbiased estimator is just the shot noise $\mse{\overline{R}} = \var{\hat{R}}/N_s$, thus we can define the signal-to-noise ratio (SNR) of the unbiased shadow mean estimator as:
\begin{align}\label{eqn:unbias_snr}
    \beta = \frac{\expect{\overline{R}}^2}{\mse{\overline{R}}} =  \frac{\expect{\hat{R}}^2}{\var{\hat{R}}/N_s}.
\end{align}
Note that $\beta^{-\frac{1}{2}}$ is the \emph{fractional error} of the unbiased shadow estimator.

We want to see if we can reduce this MSE by simply rescaling $\hat{R}$, which gives rise to the biased estimator $(1-\alpha) \hat{R}$. The MSE of this biased estimator after $N_s$ shots is:
\begin{align}
    &\quad \mse{(1-\alpha) \overline{R}} \nonumber\\
    &= \underbrace{\left(\expect{(1-\alpha) \hat{R}} - \expect{\hat{R}}\right)^2}_{\text{bias}} + \underbrace{\var{(1-\alpha) \hat{R}}/N_s}_{\text{variance}}\nonumber\\
    & =  \alpha ^2 \expect{\hat{R}}^2 +  (1-\alpha)^2\var{\hat{R}}/N_s\label{eqn:general_mse}.
\end{align}
To obtain the minimum MSE, we can take the derivative with respect to $\alpha$ and setting it to zero:
\begin{align}
   &2 \alpha^*  \expect{\hat{R}}^2 -  2(1-\alpha^*)\var{\hat{R}}/N_s = 0 \nonumber\\
   &\alpha^* = \frac{\var{\hat{R}}/N_s}{\expect{\hat{R}}^2 + \var{\hat{R}}/N_s} =
 \frac{1}{1 + \beta}. \label{eqn:alphastar}
\end{align}

The MSE at this optimal point can be obtained by substituting the expression of $\alpha^*$ back into the MSE expression, which gives
\begin{align}
    \mse{(1 - \alpha^*) \overline{R}} 
    & = \frac{\expect{\hat{R}}^2\var{\hat{R}}/N_s }{\expect{\hat{R}}^2 + \var{\hat{R}}/N_s} = \alpha^* \expect{\hat{R}}^2 \label{eqn:biased_mse}.
\end{align}
Using this, we can obtain the SNR of the optimal biased shadow mean estimator:
\begin{align}
     \beta_{\mathrm{biased}} = \frac{\expect{\hat{R}}^2}{\mse{(1 - \alpha^*) \overline{R}}} = \alpha^{*-1} = 1+\beta. \label{eqn:fractional_error}
\end{align}
This is always larger than $1$ regardless of how small the unbiased SNR $\beta$ is, i.e. using the optimal bias scheme, we can always obtain an SNR larger than $1$ for \emph{all estimators}! This is definitely not true for the unbiased estimator, e.g. in the case of local Clifford shadows that we will see later, the unbiased SNR $\beta$ decays \emph{exponentially} with the weight of the observable. 

For estimators that have large unbiased SNR $\beta \gg 1$, the improvement of the SNR from $\beta$ to $1+\beta$ by biasing is marginal. While for unbiased estimators that contain mostly noise $\beta \ll 1$, optimal biasing can always extract useful signal by increasing the SNR to above $1$. 

\subsection{Local Clifford Shadow}
In the context of shadow estimation, we are predicting the expected value of a weight-$w_p$ observable $O$ from shadows as the mean of the random variable
\begin{equation}
    \hat{R} = \tr[O\hat{\rho}].
\end{equation}
Here $\hat{\rho}$ is a snapshot of the shadow protocol.

More exactly, in the post-processing step, we prepare some stabiliser state based on the shadow and then measure the Pauli observable $O$ to obtain some \emph{random variable} $\hat{Q}$, then we rescale the result by a factor of $3^w$ to obtain the \emph{biased-free} shadow estimator $\hat{R} = 3^w \hat{Q}$. In local Clifford shadow procedure, the probability that $O$ coincides with one of the stabilisers modulo phase (the measurement basis of the shadow) is $3^{-w}$. In this case, the output from measuring $O$ on the stabiliser state is simply a Pauli random variable $\hat{O}$ with outcomes $\pm 1$, which means that $\hat{O}^2 = 1$. For the rest of the time, $O$ is not in the stabilisers modulo phase, and the output from measuring $O$ on the stabiliser state is simply $0$. Hence, we have
\begin{align*}
    \expect{\hat{Q}} &= 3^{-w}\expect{\hat{O}} + (1-3^{-w}) 0 = 3^{-w}\expect{\hat{O}},\\
    \expect{\hat{Q}^2} &= 3^{-w}\expect{\hat{O}^2} + (1-3^{-w}) 0^2 = 3^{-w},\\
    \var{\hat{Q}} &=\expect{\hat{Q}^2} - \expect{\hat{Q}}^2 =  3^{-w}  - 3^{-2w}\expect{\hat{O}}^2.
\end{align*}
Using $\hat{R} = 3^w \hat{Q}$, we have
\begin{align}
    \expect{\hat{R}} &= 3^w \expect{\hat{Q}} =  \expect{\hat{O}}\label{eqn:unbiased_est},\\
    \var{\hat{R}} &= 3^{2w} \var{\hat{Q}} =  3^{w}  - \expect{\hat{R}}^2\label{eqn:shadow_var},
\end{align}
\cref{eqn:unbiased_est} should not come as a surprise because both $\hat{R}$ and $\hat{O}$ are unbiased estimators of our target expectation value, it is just that $\hat{R}$ is obtained from the shadow procedure while $\hat{O}$ is obtain from direct measurement on the state. 

In this way, we can simplify the unbiased shadow SNR in \cref{eqn:unbias_snr} to:
\begin{align}
    \beta &= \frac{\expect{\hat{R}}^2}{\mse{\overline{R}}} =  \frac{N_s\expect{\hat{R}}^2}{3^{w}  - \expect{\hat{R}}^2} = N_s \left(\frac{3^w}{\expect{\hat{R}}^2} - 1\right)^{-1},\label{eqn:beta}
\end{align}
which fits our intuition that the SNR will increase as we increase the number of shots $N_s$, decrease the observable weight $w_p$ and/or have large expectation value $\expect{\hat{R}}$.

\subsection{Critical Biasing}
For the biased shadow estimator to outperform the unbiased one, we have:
\begin{align*}
    \mse{(1 - \alpha) \bar{R}} &\leq \mse{\bar{R}}\\
    \alpha^2 \expect{\hat{R}}^2 +  (1-\alpha)^2\var{\hat{R}}/N_s &\leq \var{\hat{R}}/N_s\\
    \alpha^2 \expect{\hat{R}}^2   &\leq \alpha(2-\alpha)\var{\hat{R}}/N_s.
\end{align*}
One trivial solution here is $\alpha = 0$. If we focus on the $\alpha > 0$ case, we then have:
\begin{align*}
    \alpha \expect{\hat{R}}^2   &\leq (2-\alpha)\var{\hat{R}}/N_s\\
    \alpha    &\leq \frac{2\var{\hat{R}}/N_s}{\expect{\hat{R}}^2 + \var{\hat{R}}/N_s} = \alpha_c.
\end{align*}
We note that $\alpha_c = 2 \alpha^*$. All the simplified expression of $\alpha^*$ using $\hat{R} = 3^w \hat{Q}$ will also apply to $\alpha_c$.

\section{\label{app:application} Possible application scenarios for biased shadow}

Here let us outline a situation where biasing might help. Suppose we try to estimate the energy $E$ of a quantum state by measuring some low-weight observable $A$ on the shadow of the state. The expectation value of $A$ can be estimated using the shadows from the estimator $\overline{A}$, and we have measure enough shadow shots such that $\var{\overline{A}}$ is small. At a later stage, we suddenly realise that there should be a higher-order correction in the form of high-weight observable $R$ with the corresponding estimator from our existing shadow denoted as $\overline{R}$. i.e. we have
\begin{align*}
    E = \expect{\overline{A}} + \expect{\overline{R}}.
\end{align*}
However, due to the high weight of the high-order correction, its variance (noise) exceed the amount of signal the it contains, i.e. its SNR is smaller than $1$:
\begin{align*}
    \expect{\overline{R}}^2/\var{\overline{R}} = \beta \leq 1.
\end{align*}
This implies that without biasing, including this higher order correction $\overline{R}$ will include more noise than signal. Hence, we are better off simply using $\overline{A}$ as our estimator for $E$ without including $\overline{R}$. The corresponding MSE of using $\overline{A}$ to estimate $E$ is
\begin{align*}
    \mse{\overline{A}}_{E} =  \var{\overline{A}} + \underbrace{\expect{\overline{R}}^2}_{\text{bias}}.
\end{align*}
As mentioned, we can increase the SNR for any estimator to above $1$ by biasing it. If we use the optimal biasing scheme for the estimation of $\expect{\overline{R}}$, this will in turn give us the following estimator for $E$:
\begin{align*}
    \overline{H} =  \overline{A} + (1-\alpha^*)\overline{R}.
\end{align*}
Using \cref{eqn:biased_mse}, we have
\begin{align*}
    \mse{\overline{H}}_{E} &=  \var{\overline{A}} + \mse{(1-\alpha^*)\overline{R}} \\
    &= \var{\overline{A}} + \alpha^* \expect{\overline{R}}^2.
\end{align*}
As mentioned before, for the original estimator $\overline{A}$ we usually have enough shots such that $\var{\overline{A}}$ is very small, thus it is likely to be the case that the higher-order correction is the error bottleneck. In this case, we can simplify the MSE of the two estimators of $E$ to
\begin{align*}
    \mse{\overline{A}}_{E} &= \expect{\overline{R}}^2\\
    \mse{\overline{H}}_{E} &= \alpha^* \expect{\overline{R}}^2.
\end{align*}
i.e. through biasing the higher-order correction, we can include it and reduce the MSE by a factor of $\alpha^{* -1} = 1+\beta$. As mentioned, the SNR of the higher-order correction was $\beta \leq 1$. Hence, we can achieve up to $2$ times reduction in the MSE in this case.

\end{document}